\documentclass[aps,prl,floatfix,showpacs,twocolumn]{revtex4}
\usepackage{amsmath,amssymb}
\usepackage{epsfig}

\begin{document}

\hsize\textwidth\columnwidth\hsize\csname@twocolumnfalse\endcsname

\title{Electron-Electron Interactions in Graphene}

\author{S. Das Sarma}
\author{Ben Yu-Kuang Hu}
\author{E.H. Hwang}
\author{Wang-Kong Tse}
\affiliation{Condensed Matter Theory Center, Department of Physics,
University of Maryland, College Park, Maryland 20742}

\begin{abstract}
We discuss the validity (or not) of the ring-diagram approximation (i.e. RPA) in the calculation of graphene self-energy in the weak-coupling ($r_s \ll 1$) limit, showing that RPA is a controlled and valid approximation for \textit{extrinsic} graphene where the Fermi level is away from the Dirac point.
\end{abstract}

\pacs{71.18.+y, 71.10.-w, 73.63.Bd, 81.05.Uw}

\maketitle
\newpage

We point out in this Comment that a recent Letter \cite{Mishch} by Mishchenko is both incorrect and misleading in asserting that the RPA self-energy is not a controlled approximation for graphene in spite of graphene being a weak-coupling two-dimensional (2D) system in the sense that the interaction coupling constant $r_s = e^2/(\hbar\kappa v)$, where $\kappa$, $v$ are respectively the background lattice dielectric constant ($\kappa \simeq 2.5$ for graphene on SiO$_2$) and the graphene band velocity ($v \approx 10^6 \mathrm{m/s}$), is small (i.e. $r_s < 1$) in graphene. We emphasize here that for \textit{extrinsic} graphene (i.e. gated or doped graphene with a free carrier density-induced chemical potential or Fermi level $\varepsilon_F$ in the conduction or valence band away from the Dirac point) the RPA is an excellent and controlled approximation which is asymptotically exact in the $r_s \to 0$ limit precisely as it is in a 2D or 3D electron gas in a jellium background \cite{Fetter,Mahan}. In \textit{intrinsic} graphene, however, where the system is a zero-gap intrinsic semiconductor with the chemical potential precisely at the Dirac point \cite{Compress}, the RPA self-energy calculation, for any arbitrary coupling-constant value (including $r_s \ll 1$), is known \cite{gpFL,Guinea} to lead to the failure of the Fermi liquid theory with the interacting system becoming a marginal Fermi liquid with a logarithmically divergent quasiparticle renormalization factor, and as such all perturbative many-body approximations become suspect, including the RPA. The sweeping statement made in Ref.~\cite{Mishch} about the lack of validity of RPA in calculating the graphene self-energy, even in the weak-coupling $r_s \ll 1$ regime, is thus incorrect for extrinsic graphene and trivial for intrinsic graphene. The conceptual confusion in Ref.~\cite{Mishch} arises from an inability to distinguish between the large-momenta ultraviolet divergence and the low-momenta infrared divergence as described below. 

In Fig.~1 we show the leading-order many-body perturbative Feynman diagrams to the graphene self-energy with the bare electron-electron interaction being the long-range Coulomb interaction $V_c(q) = 2\pi e^2/(\kappa q)$. The first-order (in $V_c$) diagrams, given by Figs.~1(a) and 1(b), are simple: The Hartree tadpole diagram (Fig.~1(a)) vanishes by virtue of the charge neutrality due to the lattice background and the Fock exchange diagram, Fig.~1(b), has been calculated and discussed in details in the literature \cite{Compress}. Of the three diagrams in the second order, only Fig.~1(d) and (e) are non-zero since the contribution of Fig.~1(c) can be trivially subsumed into the exchange self-energy diagram of Fig.~1(b). 
\begin{figure}
\includegraphics[width = 7cm]{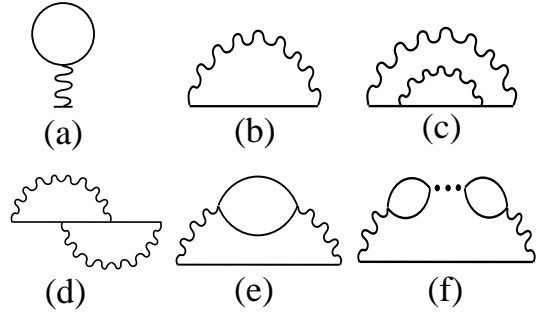} 
\caption{First-order [1(a)-1(b)] and second-order [1(c)-1(e)] self-energy diagrams. Fig.~1(f) shows the self-energy in the infinite ring-diagram approximation (i.e. RPA). The solid line denotes the electron Green function and the wavy line the bare Coulomb interaction.}
\end{figure}
For the usual parabolic dispersion jellium background electron gas problem discussed extensively in the literature \cite{Fetter,Mahan}, the long wavelength infrared (i.e. $q \to 0$) singularity of the Coulomb interaction dominates the energetics in the $r_s \ll 1$ limit, and it is well-known that the ring diagram of Fig.~1(e) has a logarithmic divergence which is fixed by summing the subset of higher-order diagrams shown in Fig.~1(f) which is the infinite series of ring diagrams. This is the RPA technique which is equivalent to keeping the leading-order term in the renormalized expansion using the effective dynamically screened Coulomb interaction $u(q) = V_c(q)/\epsilon(q,\omega)$ where $\epsilon(q,\omega) = 1-V_c(q)\Pi(q,\omega)$, with $\Pi$ being the bare electron-hole ring or bubble diagram or equivalently the non-interacting irreducible polarizability function. Note that $\epsilon$ arises from the infinite geometric series of keeping all the bubble diagrams [Fig.~1(f)]. The RPA re-summation technique completely regularizes the Coulomb infrared divergence problem and is therefore asymptotically exact in the $r_s \to 0$ limit.

Graphene, in addition to having the Coulomb infrared divergence, also has an ultraviolet divergence (i.e. $q \to \infty$) arising from its peculiar chiral linear band dispersion. This ultraviolet divergence is usually regularized by introducing a physical momentum cut-off $q_c \sim a^{-1}$ where $a$ is the graphene lattice constant. This typically introduces logarithmic terms in the cut-off momentum. This ultraviolet divergence exists for \textit{all} the self-energy diagrams for graphene, and there is nothing special about the ring diagrams as far as the ultraviolet divergence is concerned in contrast to the infrared divergence where the ring diagrams dominate in the $r_s \ll 1$ limit. For the ultraviolet divergence itself, the ring diagram [Fig.~1(e)] is not in any sense more important (or, parametrically larger) than the vertex diagram of Fig.~1(d) for any value of $r_s$.

The mathematical point for graphene, erroneously ignored in Ref.~\cite{Mishch}, is that while intrinsic graphene has only the ultraviolet divergence (where the ring diagrams and RPA have no special significance), extrinsic graphene has both the infrared and the ultraviolet divergence, making RPA a well-controlled approximation in the $r_s \ll 1$ regime. Thus, RPA is a perfectly meaningful approximation for extrinsic (i.e. doped or gated) graphene where much of the experimental work is being done.

To demonstrate the subtle mathematical point, we consider the potentially singular contribution ${\Sigma}(k,\omega)$ to the graphene self-energy coming from the ring diagram of Fig.~1(e), finding it to go as ${\Sigma}(k,\omega) = r_s^2 \int_0^{\infty}\mathrm{d}q I(q,k,\omega)/q$ where, for intrinsic graphene, $I(q,k,\omega) \sim q$ as $q \to 0$ and $\sim$ constant as $q \to \infty$. Thus intrinsic graphene has only the ultraviolet, but not the infrared, divergence, and ${\Sigma}(k) \sim \mathrm{ln}(q_c/k)$, implying that ring diagrams or RPA have no special significance for intrinsic graphene. For extrinsic graphene, however, $I(q) \sim$ constant for both $q \to 0$ and $q \to \infty$, and therefore, the infrared (as well as the ultraviolet) divergence is present, making RPA a meaningful well-controlled approximation \cite{Hwang,Polini}, as in ordinary electron gas systems in the weak-coupling regime. In extrinsic graphene, which is the experimentally relevant graphene system with free carriers, RPA regularizes the usual infrared divergence associated with the long-range Coulomb interaction whereas RPA is not meaningful in intrinsic graphene which does not manifest any infrared Coulomb divergence. More details are given in the appendix. 

We point out that in graphene, even extrinsic graphene where the carrier density can be varied, the coupling parameter $r_s = e^2/(\hbar \kappa v)$ does not depend on the density, in sharp contrast to the usual 2D ($r_s \sim n^{-1/2}$) and 3D ($r_s \sim n^{-1/3}$) parabolic dispersion electron systems. Thus, in graphene $r_s$ ranges between $2.4$ (for free standing graphene with $\kappa = 1$) and $0$ --- e.g. $r_s \simeq 0.9$ for graphene on SiO$_2$, $r_s \simeq 0.15$ for graphene on HfO$_2$. Therefore, in some sense RPA is more accurate in extrinsic graphene than in 3D metals ($r_s \approx 3-6$) since extrinsic graphene (with $r_s \approx 0.15-2.4$) better satisfies the weak-coupling ($r_s < 1$) condition necessary for the quantitative validity of RPA.

Finally, we also mention that the main result reported in Ref.~\cite{Mishch}, namely that the
intrinsic conductivity in graphene is renormalized by electron-electron
interactions, is manifestly incorrect as it violates the well-known Kohn
theorem. The long wavelength conductivity cannot be modified by 
electron-electron interaction in a translationally invariant system by
virtue of the separation of center of mass and relative coordinates, and the
vertex correction and the self-energy diagrams must cancel each other in
each order, as shown recently explicitly for graphene in Ref.~\cite{Sheehy}.

\section{Appendix}

The kinetic energy of graphene for 2D wave vector {\bf
  k} is given by
$\epsilon_{s{\bf k}} = s v |{\bf k}|$,
(we use $\hbar = 1$ throughout this paper)
where $s=\pm 1$ indicate the conduction (+1) and valence ($-1$) bands,
respectively, and $v$ is the Fermi velocity. 
The corresponding density of states (DOS) is given by
$ D(\varepsilon) = g |\varepsilon|/(2\pi v^2)$, where
$g=g_s g_v$ with $g_s=2$, $g_v=2$ the spin and valley degeneracies, respectively. 
The Fermi momentum $k_F$ and the Fermi energy $\varepsilon_F$ of 2D graphene are 
given by $k_F = (4\pi n/g_s g_v)^{1/2}$ and $\varepsilon_F = vk_F$ where 
$n$ is the 2D carrier (electron or hole) density. For intrinsic (extrinsic) graphene, 
$n$, $\varepsilon_F =(\neq) 0$ with the Dirac point taken to be the energy zero.

The exchange energy graph is the first-order diagram of interaction
shown as in Fig. 1(b). It 
is the only contribution with one Coulomb line. 
The exchange self-energy is given by
\begin{equation}
\Sigma^{\rm ex}_{s}({\bf k}) = -\sum_{s'{\bf q}} 
\ n_F(\xi_{{\bf k}+{\bf q}s'})\, V_c({\bf q})\, F^{(1)}_{ss'}({\bf
k}, {\bf k}+{\bf q}),
\label{eq:1} 
\end{equation}
where $\xi_{{\bf k}s} = \epsilon_{s{\bf k}} - \varepsilon_F$, $n_F(\xi_{{\bf
    k}s}) = \theta(k_F-s\epsilon_{\bf k})$ is the Fermi 
function at $T=0$, $F^{(1)}_{ss'}({\bf k},{\bf k}+{\bf q})$ is
the band overlap matrix element given by
\begin{equation}
F^{(1)}_{ss'}({\bf k},{\bf k}') = \frac{1}{2}(1 + ss' \cos\theta_{{\bf
    kk}'}), 
\end{equation}
where $\theta_{{\bf kk}'}$ is the
angle between ${\bf k}$ and ${\bf k}'$.

There are three diagrams in the second-order self-energy diagrams with two Coulomb lines.
Fig. 1(d) can be expressed as
\begin{widetext} 
\begin{equation}
\Sigma_s^{(d)}({\bf k},ik_n)  = 
\frac{1}{\beta^2}\sum_{s_1,s_2,s_3}
\sum_{\bf q,q'} \sum_{iq_n,iq_n'}V_c({\bf q})V_c({\bf q}') 
F^{(2)}_{ss_1s_2s_3}({\bf k},{\bf k}_1,{\bf k}_2,{\bf k}_3)
G^0_{s_1}({\bf k}_1,ik_{1n}) G^0_{s_2}({\bf k}_2,ik_{2n}) G^0_{s_3}({\bf k}_3,ik_{3n}),
\end{equation}
where $\beta = 1/k_B T$, $s,s_i=\pm 1$ denote the band indices, 
$G_s^0({\bf k},ik_n)=1/(ik_n-\xi_{{\bf k}s})$ is the
unperturbed Green function, $V_{c}(q)=2\pi e^2/\kappa q$ is the bare Coulomb potential (with
background dielectric constant $\kappa$). And
the function $F^{(2)}$ is the band overlap matrix element, given by
\begin{equation}
F^{(2)}_{s_1s_2s_3s_4}({\bf k}_1,{\bf k}_2,{\bf k}_3,{\bf k}_4) =
\frac{1}{8}
\left [ 1+ s_1s_2s_3s_4 + \frac{1}{2} \sum_{i \neq j}^4 s_i s_j \cos
  \theta_{{\bf k}_i,{\bf k}_j} \right ].
\end{equation}
After summing over frequencies and a standard procedure of analytical
continuation,  we have 
\begin{eqnarray}
\Sigma_s^{(d)}({\bf k},\omega)  & = & -\sum_{s_1,s_2,s_3}
\sum_{\bf q,q'} \frac{V_c({\bf q})V_c({\bf q}')}{\omega+i\delta -\xi_{{\bf
      k}_1s_1} + \xi_{{\bf k}_2s_2} - \xi_{{\bf k}_3s_3} }
F^{(2)}_{ss_1s_2s_3}({\bf k},{\bf k}_1,{\bf k}_2,{\bf k}_3) \nonumber \\
&\times& \left \{ n_F(\xi_{{\bf k}_3 s_3}) \left [ n_F(\xi_{{\bf k}_1
      s1}) - n_F(\xi_{{\bf k}_2s_2}) \right ] + 
n_F(\xi_{{\bf k}_2 s_2}) \left [ 1- n_F(\xi_{{\bf k}_1 s_1}) \right ]
\right \} \nonumber \\
&=& \sum_{s_1,s_2,s_3}\sum_{\bf q,q'} R_{ss_1s_2s_3}^{(d)}({\bf k},\omega,{\bf q,q'}).
\end{eqnarray}
\end{widetext}
For intrinsic graphene $n_F(\xi_{{\bf k}+}) = 0$ and $n_F(\xi_{{\bf
    k}-}) = 1$. Thus, only the following terms in $\Sigma_{+}^{(d)}$ are
nonzero for intrinsic graphene: 
$R_{++-+}^{(d)}$ and $R_{+-+-}^{(d)}$. As $q \rightarrow \infty$,
these functions behave as $1/q$. 
Therefore the self-energy contributions by
these terms are logarithmically divergent. 
After introducing a physical cutoff ($q_c$) we have $\Sigma_s^{(d)} \propto
C_d \ln (q_c/k)$, where $C_d$ is a regular function of $k,\omega$. As
$q \rightarrow 0$ the integrands $R_{++-+}$ and $R_{+-+-}$ are
finite and introduce no singular behaviors in $\Sigma_s^{(d)}$. 

For extrinsic graphene $n_F(\xi_{{\bf k}+}) = \theta (k_F - |{\bf k}|)$
and $n_F(\xi_{{\bf k}-}) = 1$. Thus, we have five non-zero
terms in Eq. (5):
$R_{++++}^{(d)}$, $R_{+++-}^{(d)}$, $R_{++-+}^{(d)}$,
$R_{+-++}^{(d)}$, and $R_{+-+-}^{(d)}$.
Here $R_{++++}^{(d)}$, $R_{+++-}^{(d)}$, and 
$R_{+-++}^{(d)}$ do not show any singular behavior in both $q
\rightarrow 0$ and $q \rightarrow \infty$ limits. Again
$R_{++-+}^{(d)}$ and $R_{+-+-}^{(d)}$ show 
singular behavior, $\propto 1/q$, only as $q \rightarrow \infty$. Thus we
can control this singularity by introducing a physical cutoff.
In summary of $\Sigma_{s}^{(d)}$, we have an ultraviolet divergence in
the integrand both for intrinsic and extrinsic graphene. However, we
can remove this divergence by introducing a cutoff $q_c \sim 1/a$.

Fig. 1(e) can be written as
\begin{widetext}
\begin{equation}
\Sigma_s^{(e)}({\bf k},ik_n)  = 
\frac{1}{\beta}\sum_{s_1}
\sum_{\bf q} \sum_{iq_n}[V_c({\bf q})]^2
F^{(1)}_{ss_1}({\bf k},{\bf k + q}) \Pi({\bf q},iq_n)
G^0_{s_1}({\bf k+q},ik_n + iq_n),
\end{equation}
where $\Pi({\bf q},iq_n)$ is the polarizability (bare bubble diagram)
of graphene, given by 
\begin{equation}
\Pi({\bf q},iq_n) = g \sum_{s_1,s_2}\sum_{\bf k} \frac{n_F(\xi_{{\bf
      k}s_1}) - n_F(\xi_{{\bf k + q}s_2})} {iq_n + \xi_{{\bf k}s_1} -
  \xi_{{\bf k+q}s_2}}
F^{(1)}_{s_1s_2}({\bf k},{\bf k +  q}).
\end{equation}
We can express Eq. (6) as $\Sigma_s^{(e)}({\bf k},\omega) = \Sigma_s^{\rm
  line}({\bf k},\omega) + \Sigma_s^{\rm pole}({\bf k},\omega)$
after a standard procedure of analytical
continuation: 
\begin{equation}
\Sigma_s^{\rm line}({\bf k},\omega)  = - \sum_{s_1}
\sum_{\bf q} \int \frac{d\omega'}{2\pi} [V_c({\bf q})]^2
F^{(1)}_{ss_1}({\bf k},{\bf k + q}) \frac{\Pi({\bf
    q},i\omega')}{i\omega'+\omega-\xi_{{\bf k+q}s_1}},
\end{equation}
\begin{equation}
\Sigma^{\rm pole}_s({\bf k},\omega) = \sum_{s_1{\bf q}}\left [
  \theta(\omega - \xi_{{\bf k+q}s'})-\theta(-\xi_{{\bf k+q}s'}) \right ]
[V_c({\bf q})]^2F^{(1)}_{ss_1}({\bf k,k+q})\Pi({\bf q},\xi_{{\bf k+q}s_1}-\omega).
\end{equation}
\end{widetext}
Here $\Sigma_s^{\rm pole}$ does not show any singular behavior and is 
a well-defined function for all {\bf k} and $\omega$.
We can rewrite $\Sigma_s^{\rm line}$ as
\begin{equation}
\Sigma_s^{\rm line}({\bf k},\omega) = -r_s^2\frac{v^2}{2\pi}
\int_0^{\infty} \frac{dq}{q}I(q,k,\omega).
\end{equation}
For intrinsic graphene we have $I(q) \sim q$ as  $q
\rightarrow 0$, and $I(q) \sim $ constant as $q \rightarrow \infty$.
Thus, we have only the ultraviolet divergence in the integrand, but
no infrared divergence. Again, $\Sigma_s^{(e)} \propto \ln (q_c/k)$. 
For extrinsic graphene we have a different behavior as $q \rightarrow
0$, i.e. $I(q) \sim $ constant as $q \rightarrow 0$. 
Therefore, we have both infrared and ultraviolet divergences in the
integrand. This singular 
behavior arises from the $1/q$ nature of the Coulomb interaction. We cannot control 
this divergence by introducing a cutoff, and the $q \to 0$ divergence must be regularized by the usual
infinite ring-diagram approximation as in the usual RPA [i.e. Fig.~1(f)].  

In conclusion, we have the ultraviolet divergence controlled by a cutoff for all diagrams but 
only Fig. 1(e) (RPA-type diagram) shows the infrared divergence for
extrinsic graphene, which has to be controlled by the RPA re-summation given in Fig.~1(f).

\end{document}